\documentclass[aps,prd,12pt,preprint,tightenlines,superscriptaddress,
amsfonts,amssymb,amsmath,byrevtex,showpacs,nofootinbib]{revtex4}
\begin{document}
\newcommand{\dR}{\mathbb R}
\newcommand{\dS}{\mathbb S}
\newcommand{\id}{\mathbb I}
\newcommand{\Milne}{\mathcal{M}}
\newcommand{\sing}{\mathcal{S}}
\newcommand{\dZ}{\mathbb Z}
\newcommand{\up}{\uparrow}
\newcommand{\down}{\downarrow}
\newcommand{\var}{\varepsilon}
\newcommand{\Om}{\Omega_{\check{\theta}}}
\newcommand{\Ola}{\Omega_{\check{\lambda}}}

\title{Simple model of big-crunch/big-bang transition}

\author{Przemys\l aw Ma\l kiewicz$^\dag$ and W\l odzimierz Piechocki$^\ddag$
\\ Department of Theoretical Physics\\So\l tan Institute for Nuclear Studies,
\\ Ho\.{z}a 69, 00-681 Warszawa, Poland;
\\ $^\dag$pmalk@fuw.edu.pl, $^\ddag$piech@fuw.edu.pl}

\date{\today}
\begin{abstract}
We present classical and quantum dynamics of a test particle in the
compactified Milne space. Background spacetime includes one compact
space dimension undergoing contraction  to a point followed by
expansion.  Quantization consists in finding a self-adjoint
representation of the algebra of particle observables.  Our model
offers some insight into the nature of the cosmic singularity.
\end{abstract} 
\maketitle

\section{Introduction}

Presently available cosmological data  suggest that our universe
emerged from a state with extremally high density of physical fields
\cite{Spergel:2003cb,Bahcall:1999xn}. It is called the cosmic
singularity. For modelling the very early universe it is necessary to
understand the nature of the singularity. We  also believe that the
cosmic singularity problem is inseparable from the problem of dark
energy, which seems to be the most fundamental problem of
contemporary physics (see, e.g.
\cite{rbj,Peebles:2002gy,Padmanabhan:2002ji} and references therein).

It is attractive to assume that the singularity consists of
contraction and expansion phases. This way one opens door for the
cyclic universe models
\cite{Khoury:2001bz,Steinhardt:2001vw,Steinhardt:2001st,
Steinhardt:2004gk,Bojowald:2004kt,Gasperini:2002bn,Lidsey:1999mc},
which are expected to be free of the problem of beginning/end  of the
universe , i.e. creation/annihilation of space-time-matter-field
from/into `nothing'. It is also proposed to use such models to get
rid of the cosmic inflation \cite{Linde:2005ht,Linde:2002ws}, because
of its puzzling features and difficulties
\cite{Turok:2004yx,Turok:2002yq,GDS}.

There exist at least two frameworks  used for modelling the
singularity: general relativity (GR) and string/M (SM)
theory\footnote{Since string/M theory is far from being complete, we
make distinction between SM and GR.}. One of the simplest models of
singularity offered by GR is described by de Sitter's
space\footnote{It is only symbolically called here the singularity;
de Sitter space with topology $\dR^1\times\dS^1$ has neither
incomplete geodesics nor blowing up Riemann tensor invariants.}. It
is of the `big-bounce' type, which may be treated as a gentle
singularity. The Milne space, considered recently within SM scheme
\cite{Khoury:2001bz}, is the simplest spacetime modelling the
`big-crunch/big-bang' type singularity. It represents a violent model
of the contraction/expansion transition.

Any reasonable  model of the cosmic  singularity should be able to
describe quantum propagation of a fundamental object (e.g. particle,
string, membrane,...) from the pre-singularity to post-singularity
epoch. It is the most elementary criterion that should be satisfied.
Some insight into the problem may be already achieved  by studying
dynamics of a test particle in low dimensional spacetime. Recently,
we presented results concerning dynamics of a particle in the
two-dimensional  de Sitter space \cite{WP,Piechocki:2003hh}. The
model has passed the above test. Classical and quantum dynamics of a
particle is well defined in the entire universe including the
big-bounce period.

Results of the present paper concern evolution of a particle in the
compactified Milne space. We analyse classical and quantum dynamics
of a free particle in two-dimensional spacetime. In Section II we
specify geometry, topology and symmetry of the Milne space. Classical
dynamics of a particle is carried out in Section III. We find
constraint for dynamics, determine dynamical integrals, identify
observables and introduce phase space of the system. Next, we analyse
particle's motion in the Milne space and specify local symmetry of
the phase space. We consider four models of evolution of a particle
across the singularity. Section IV is devoted to quantization of the
classical system. Quantization is carried out by finding a
self-adjoint representation of the algebra of observables.  In
Section V we suggest the way  our  results can be linked to
cosmological solutions of some higher-dimensional effective field
theories. We conclude in Section VI.

\section{The compactified Milne space}

\subsection{Geometry and topology}

Let us define two quadrants, $\Milne $, of the two-dimensional
Minkowski space by \begin{equation}\label{Mil}
    \mathcal{M}=\{(x^+,x^-)\in\dR^2~|~x^+ x^- > 0~\vee~x^+ = 0 =x^-\},
    ~~~~~x^\pm := x^0 \pm x^1,
\end{equation}
where $x^0~$ (time) and $x^1~$ (space) are coordinates of the
Minkowski space with the signature defined by the line element
$ds^2= -(dx^0)^2 + (dx^1)^2$.  Next, let us introduce the finite
boost transformation $B$ on $\Milne$
\begin{equation}\label{boost}
    B: (x^+,x^-)\longrightarrow (e^{2\pi r} x^+, e^{-2\pi r} x^-),
\end{equation}
where $r$  defines a boost.  The factor space $\Milne/B$ is called
the compactified Milne space, $\Milne_C$. Since specification of $r$
identifies the $\Milne_C$ space, the set of  the compactified Milne
spaces is uncountable.

It is convenient to introduce new coordinates $t$ and $\theta$
defined by \begin{equation}\label{Min}
    x^0 =: t\:\cosh\theta,~~~~~~ x^1 =: t\:\sinh\theta .
    \end{equation}
The line element in  $\Milne_C$ reads
\begin{equation}\label{line}
    ds^2 =  - dt^2 +t^2 d\theta^2 ,
\end{equation}
where $(t,\theta)\in \dR^1 \times \dS^1$.

To visualize the compactified Milne space, we  present the
isometric embedding of $\Milne_C$ into three-dimensional Minkowski
space. It may be defined by the mapping
\begin{equation}\label{embed}
    y^0(t,\theta) = t\sqrt{1+r^2},~~~~ y^1(t,\theta) = rt\sin(\theta/r),
    ~~~~y^2(t,\theta) = rt\cos(\theta/r),
\end{equation}
and one has
\begin{equation}\label{con}
    \frac{r^2}{1+r^2}(y^0)^2 - (y^1)^2- (y^2)^2 =0. \end{equation}
Eq. (\ref{con}) presents two cones with a common vertex at
$\:(y^0,y^1,y^2)= (0,0,0)$. One may  verify that the induced metric
on (\ref{con}) coincides with the metric (\ref{line}).

The  space  $\Milne_C$ is locally isometric with the Minkowski space
at each point, but at the vertex $t=0$. The neighborhood of that very
special point cannot be made homeomorphic to an open circle in $\dR
^2$. For this reason $\Milne_C$ is not a manifold, but orbifold.
Obviously, the Riemann tensor is not well defined there. At the locus
$t=0$ the orbifold  $\Milne_C$ has space-like singularity. However,
it is of removable type  because any time-like geodesic from the
lower cone ($t<0$) linked with the vertex ($t=0$) can be extended to
the time-like geodesic of the upper-cone ($t>0$). It is clear that
such an extension cannot be unique because at $t=0$ the Cauchy
problem for the geodesic equation is not well defined, due to the
disappearance of space dimension.

The coefficient of $\:d\theta^2\:$ in (\ref{line}) disappears as
$t\rightarrow 0$. Therefore, one may use the $\Milne_C$ space to
model a two-dimensional universe with the `big-crunch/big-bang'
singularity.

\subsection{Local symmetry}

Solution to the Killing field equations with the metric
(\ref{line}) reads
\begin{equation}\label{Kil} \eta_1 =
\cosh\theta\: \frac{\partial}{\partial t}-
\frac{\sinh\theta}{t}\:\frac{\partial}{\partial \theta},~~~~\eta_2
= \sinh\theta \:\frac{\partial}{\partial t}
-\frac{\cosh\theta}{t}\:\frac{\partial}{\partial
\theta},~~~~\eta_3 = \frac{\partial}{\partial \theta}.
\end{equation}

One may easily verify that the Killing vectors (\ref{Kil}) satisfy
the algebra \begin{equation}\label{com1}
    [\eta_1 , \eta_2]=0,~~~~~[\eta_3 , \eta_2]=\eta_1,~~~~~[\eta_3 ,
    \eta_1]=\eta_2,
\end{equation}
which is the $iso(1,1)$ Lie algebra \cite{Vil}.   The algebra
(\ref{com1}) is well defined locally everywhere in $\Milne_C$ with
exception of the singularity $t=0$.

\section{Classical dynamics of a particle}

An action integral, $\mathcal{A}$, describing a relativistic test
particle of mass $m$ in a gravitational field $g_{kl},~~(k,l=0,1)$
may be defined by \begin{equation}\label{action} \mathcal{A}=\int
d\tau\: L(\tau),~~~~~~~ L(\tau):=\frac{m}{2}\:(\frac{\dot{x}^k
\dot{x}^l}{e} g_{kl}-e),~~~~\dot{x}^k :=dx^k/d\tau, \end{equation}
where $\tau$ is an evolution parameter, $e(\tau)$ denotes the
`einbein' on the world-line, $x^0$ and $x^1$ are time and space
coordinates, respectively.

\noindent In case of $\Milne_C$ space the Lagrangian
reads\footnote{It is not well defined for $t=0$, unless one can
give meaning to $\dot{\theta}$ at $t=0$; equations (\ref{con1})
and (\ref{var}) suffer from the same problem.}.
\begin{equation}\label{lag}
    L(\tau)=  \frac{m}{2e}\:(t^2 \dot{\theta}^2 -\dot{t}^2 -e^2).
\end{equation}
The action (\ref{action}) is invariant under reparametrization
with respect to $\tau$. This gauge symmetry leads to the constraint
\begin{equation}\label{con1}
    \Phi :=  (p_\theta/t)^2 - (p_t)^2 + m^2 =0,
\end{equation}
where  $p_t := \partial L/\partial\dot{t}\:$ and $p_\theta
:=\partial L/\partial\dot{\theta}\:$ are  canonical momenta.

Variational principle applied to (\ref{action}) gives at once the
equations of motion of a particle
\begin{equation}\label{var}
\frac{d}{d\tau}p_t - \frac{\partial L}{\partial t} = 0,~~~~~~
\frac{d}{d\tau} p_\theta = 0, ~~~~~~\frac{\partial L}{\partial e} =
0. \end{equation} Thus, during  evolution of the system $p_\theta$ is
conserved. Owing to the constraint (\ref{con1}), $p_t$ blows up as
$t\rightarrow 0~$ for $p_\theta \neq 0 $. This is a real problem,
i.e.  it cannot be avoid by a suitable choice of coordinates. It is
called the 'blue-shift' effect. However, trajectories of a test
particle, i.e. nonphysical particle, coincide (by definition) with
time-like geodesics of an empty spacetime, and there is no obstacle
for such geodesics to reach/leave the singularity\footnote{We
continue this discussion in the subsections A, B, C and D, and in the
conclusion section.}.

It is commonly known that  Killing vectors of a spacetime may be
used to find dynamical integrals of a particle, i.e. quantities
which do not change during the motion of a point mass. In our case
there exist three dynamical integrals and they can be determined
as follows
\begin{equation}\label{in1} I_1 := p_t\:\eta_1^t +
p_\theta\:\eta_1^\theta = p_t\cosh\theta -
p_\theta\frac{\sinh\theta}{t}, \end{equation}
\begin{equation}\label{in2} I_2 := p_t\:\eta_2^t +
p_\theta\:\eta_2^\theta = p_t\sinh\theta -
p_\theta\frac{\cosh\theta}{t}, \end{equation}
\begin{equation}\label{in3} I_3 := p_t\:\eta_3^t +
p_\theta\:\eta_3^\theta =   p_\theta ,
\end{equation}
where $\eta_a^T$ and  $\eta_a^\theta$ are components of the
Killing vectors $\eta_a ~(a=1,2,3)$. Making use of
(\ref{in1})-(\ref{in3}) we may rewrite the constraint (\ref{con1})
in the form \begin{equation}\label{con2}
    \Phi = I_2^2 -  I_1^2 + m^2 = 0.
\end{equation}

For further analysis we introduce the phase space. It is defined
to be the space of all particle geodesics. To describe a geodesic
uniquely one may use two independent dynamical integrals. In case
only one part of the Milne space is available for particle
dynamics, for example with $t<0$, the phase space, $\Gamma$, could
be defined as
\begin{equation}\label{phase}  \Gamma =
\{(I_1,I_2,I_3)~|~I_2^2 - I_1^2 + m^2 =0, \: I_3 = p_\sigma \}.
\end{equation}
For the choice  (\ref{phase}) the phase space may be parametrized by
two variables $\sigma$ and $p_\sigma$ in the following way
\begin{equation}\label{par} I_1 = m\:\cosh\sigma,~~~~~I_2 = m\:
\sinh\sigma,~~~~~I_3 = p_\sigma.
\end{equation} One can easily check that
\begin{equation}\label{coom}
\{I_1,I_2\}=0,~~~~~\{I_3,I_2\}=I_1,~~~~~\{I_3,I_1\}=I_2,
\end{equation}
where the Poisson bracket is defined as
\begin{equation}\label{bra1}
\{\cdot,\cdot\} = \frac{\partial \cdot}{\partial
p_\sigma}\frac{\partial\cdot}{\partial\sigma} -
\frac{\partial\cdot}{\partial \sigma}\frac{\partial\cdot}{\partial
p_\sigma}.
\end{equation} Thus the dynamical integrals (\ref{in1})-(\ref{in3})
and the Killing vectors (\ref{Kil}) satisfy the same algebra. Using
properties of the Poisson bracket we get \begin{equation}\label{com2}
\{\Phi,I_a\} = 0,~~~~~a=1,2,3.
\end{equation}

 We define classical observables to be real functions on phase space
which are: (i) gauge invariant,  (ii) specify  all time-like
geodesics of a particle, and (iii) their algebra corresponds to the
local symmetry of the phase space. It is clear, due to (\ref{con2})
and (\ref{com2}), that all dynamical integrals are gauge invariant.
There exist two  functionally independent combinations of them which
specify all  time-like geodesics (see (\ref{r4}) of appendix). We use
them to represent particle observables (one may verify that they are
gauge invariant).

\subsection{Specification of phase space and observables based on continuous symmetries}

Let us denote by $\sing_\down$ the part of spacetime $\Milne_C$
with $t<0$, the big-crunch/big-bang singularity by $\sing$, and
the part of $\Milne_C$ with $t>0$  by   $\sing_\up$.

By definition, a test particle with constant mass does not modify a
background spacetime. Hence, we postulate that a particle arriving at
the singularity $\sing$ from $\sing_\down$ is `annihilated' at
$\sing$ and next, `created' into $\sing_\up$. There are four
interesting cases of propagation  depending on the way a particle may
go across $\sing$. In each case the propagation must be consistent
with the constraint equation (\ref{con1}) and have constant
$p_\theta$ (due to (\ref{var})) in $\sing_\down$ and $\sing_\up$. At
$\sing$ both (\ref{con1}) and (\ref{var}) are not well defined.

In this subsection we consider the following propagation:
 particle following spiral geodesics winding clockwise the cone
$\sing_\down $ continues to move along clockwise spirals in
$\sing_\up$ (the same concerns propagation along anticlockwise
spirals). Obviously, for $p_\theta =0$ particle trajectories are just
straight lines both in $\sing_\down$ and $\sing_\up$. Apart from this
we take into account the rotational invariance (with respect to the
axis which coincides with the $y^0$-axis of 3d Minkowski frame
defining (\ref{embed})) of the space of particle trajectories which
occur independently in $\sing_\down$ and $\sing_\up$.

It results from  (\ref{soll}) of Appendix A that the set of all
particle trajectories can be  determined by two parameters
$\:(c_1,c_2)\in \dR^1 \times [0,2\pi[$. Thus, the phase space
$\Gamma_\down$ of a particle in $\sing_\down$ has topology $\dR^1
\times \dS^1$. The transition of a particle across $\sing$ makes
the dynamics in $\sing_\down$ and $\sing_\up$ to be, to some
extent, independent so the phase space $\Gamma_\up$ of a particle
in $\sing_\up$ has also the  $\dR^1 \times \dS^1$ topology.
Therefore, the phase space $\Gamma_C$ of the entire system has the
topology $\dS^1\times\dR^1\times\dS^1$.

Now let us specify the local symmetry of either $\Gamma_\down$ or
$\Gamma_\up$ by defining the Lie algebra of particle observables.
The system has two independent degrees of freedom represented by
the observables $c_1$ and $c_2$. Equation (\ref{soll}) tells us
that $c_2$ has interpretation of position coordinate, whereas
$c_1$ plays the role of momentum, owing to (\ref{r3}). With such
an interpretation, it is natural to postulate the following Lie
algebra for either $\Gamma_\down$ or $\Gamma_\up$.
\begin{equation}\label{com3} \{c_1,c_2\}=1,~~~~~~ \{\cdot,\cdot\}
:= \frac{\partial\cdot}{\partial c_1}\frac{\partial\cdot}{\partial
c_2} - \frac{\partial\cdot}{\partial
c_2}\frac{\partial\cdot}{\partial c_1}.
\end{equation} Heuristic reasoning we use to introduce the algebra
(\ref{com3}) may be replaced by  derivation. It is presented at
the end of Appendix A.

Suppose the observables $c_1$ and $c_2$ describe dynamics in
$\sing_\down$, and let us assume that propagations  in
$\sing_\down$ and $\sing_\up$ are independent. In such case it
would be convenient to introduce  two new observables $c_4$ and
$c_3$ in $\sing_\up$ corresponding to $c_1$ and $c_2$. The Lie
algebra in $\Gamma_C$ would be defined as follows
\begin{equation}\label{com4}
\{c_1,c_2\}=1,~~~~\{c_4,c_3\}=1,~~~~\{c_i,c_j\}=0,~~~~~
\textrm{where}~~i=1,2~~~\textrm{and}~~~j=3,4
\end{equation}
with the Poisson bracket
\begin{equation}\label{com5}
\{\cdot,\cdot\} := \frac{\partial\cdot}{\partial
c_1}\frac{\partial\cdot}{\partial c_2} +
\frac{\partial\cdot}{\partial c_4}\frac{\partial\cdot}{\partial c_3}
- \frac{\partial\cdot}{\partial c_2}\frac{\partial\cdot}{\partial
c_1} - \frac{\partial\cdot}{\partial
c_3}\frac{\partial\cdot}{\partial c_4}.
\end{equation} But from the discussion above it results that
$\Gamma_C$ has only three independent variables.  We can encode this
property modifying (\ref{com4}) and (\ref{com5}) by the condition
$c_4= c_1$. Finally, we get \begin{equation}\label{com6}
\{c_1,c_2\}=1,~~~~~ \{c_1,c_3\}=1,~~~~ \{c_2,c_3\}=0, \end{equation}
with the Poisson bracket \begin{equation}\label{com7} \{\cdot,\cdot\}
= \frac{\partial\cdot}{\partial c_1}\frac{\partial\cdot}{\partial
c_2} + \frac{\partial\cdot}{\partial
c_1}\frac{\partial\cdot}{\partial c_3} -
\frac{\partial\cdot}{\partial c_2}\frac{\partial\cdot}{\partial c_1}
- \frac{\partial\cdot}{\partial c_3}\frac{\partial\cdot}{\partial
c_1}.
\end{equation}

The  type of propagation we have considered so far is consistent
with the isometry (i.e., continuous symmetry) of the compactified
Milne space. In the next subsection we increase respected
symmetries to include the space inversion (i.e., discrete
symmetry).

\subsection{Specification based on continuous and discrete symmetries}

We take into account (as in  case considered in subsection A) that
$\sing_\down$ and $\sing_\up$ have the (clockwise and
anticlockwise) rotational symmetry quite independently. Apart from
this we assume that the singularity $\sing$ may `change' the
clockwise type geodesics into anticlockwise ones, and
\textit{vice-versa}.  From mathematical point of view such case is
allowed because at $\sing$ the space dimension disappears, thus
$p_\theta$ is not well defined there, so it may have different
signs in $\sing_\down$ and $\sing_\up$. Therefore, the space of
geodesics has reflection type of symmetry independently in
$\sing_\down$ and $\sing_\up$, which is equivalent to the space
inversion separately in $\sing_\down$ and $\sing_\up$. The last
symmetry is of discrete type, so it is not the isometry of the
compactified Milne space. It is clear that the phase space
$\Gamma_C$ has the topology $\dS^1\times\dR^1\times\dS^1\times
\dZ_2$.

Proposed type of propagation of a particle through $\sing$ may be
characterized by the conservation of $|p_\theta|$ (instead of
$p_\theta $  required in  subsection A). The consequence is that now
$|c_1|=|c_4|$ (instead of $c_1=c_4$ of subsection A). To obtain the
algebra of observables we propose to put $c_4=\varepsilon c_1$, where
$\varepsilon =\pm 1$ is a new descrete variable, into (\ref{com4})
and (\ref{com5}). Thus the algebra reads
\begin{equation}\label{com8}
\{c_1,c_2\}=1,~~~~~
\{c_1,c_3\}=\varepsilon,~~~~ \{c_2,c_3\}=0,
\end{equation} with
the Poisson bracket
\begin{equation}\label{com9} \{\cdot,\cdot\} =
\frac{\partial\cdot}{\partial c_1}\frac{\partial\cdot}{\partial c_2}
+ \varepsilon\frac{\partial\cdot}{\partial c_1}\frac{\partial\cdot}
{\partial c_3} - \frac{\partial\cdot}{\partial
c_2}\frac{\partial\cdot}{\partial c_1}
 - \varepsilon\frac{\partial\cdot}{\partial c_3}\frac{\partial\cdot}  {\partial c_1}.
 \end{equation}

\subsection{The case trajectories in pre- and post-singularity epochs are independent}

Now, we assume that there is no connection at all between
trajectories in the upper and lower parts of the Milne space. For
instance, spiral type geodesic winding  the cone in $\sing_\down $
may be `turned' by $\sing$ into straight line in  $\sing_\up $, and
\textit{vice-versa}. It means that we consider the case equations
(\ref{con1}) and (\ref{var}) are satisfied both in $\sing_\down $ and
$\sing_\up $, but not necessarily at $\sing$ (as in case considered
in subsection B). In addition we propose that $p_\theta$ may equal
zero either in $\sing_\down $ or in $\sing_\up $. Justification for
such choices are the same as in the preceding subsection. Obviously,
the present case also includes transitions of spiral geodesics into
spiral ones, and straight line into straight line geodesics.

It is clear that now  the algebra of observables coincides with
(\ref{com4}) and (\ref{com5}), and the entire phase space
$\Gamma_C$ has  the topology $\Gamma_\down\times\Gamma_\up :=
(\dS^1\times\dR^1)\times(\dR^1\times\dS^1)$.

\subsection{The case space of trajectories has reduced form of rotational invariance}

There is one more case  we would like to consider: it can be obtain
from the case considered in subsection A by ignoring the rotational
invariance of the space of solutions assumed to exist separately in
$\sing_\down $ and $\sing_\up $. Now we assume that the invariance
does occur, but in the entire spacetime. Consequently, the algebra of
observables is defined by Eq. (\ref{com3}).

Such type of symmetry of the space of geodesics appears,  e.g.  in
case of propagation of a particle in two-dimensional one-sheet
hyperboloid embedded in three-dimensional Minkowski space
\cite{WP} (2d de Sitter space with topology $\dR^1 \times \dS^1$).

\section{Quantization}

By quantization we mean finding a self-adjoint representation of the
algebra of classical observables\footnote{We do not need the
observables to be well defined globally, which would be required for
finding an unitary representation of the corresponding Lie group.}.
We find that our quantization method  is sufficient for analysis of
evolution of a quantum particle across the vertex of $\Milne_C$. Such
method was used in our previous papers \cite{WP,Piechocki:2003hh}
dealing with dynamics of a particle in de Sitter
space\footnote{Lifting of self-adjoint representation of the algebra
to the unitary representation of the corresponding Lie group was
possible in case of the spacetime topology $\dR^1 \times \dS^1$, but
could not be done in case of topology $\dR^2$.}. Applying the same
quantization method in both cases enables the comparison of results.

In this paper our genuine spacetime is $\Milne_C$, i.e. the Milne
space $\Milne$ is only used as a tool for defining $\Milne_C$. Thus
our main concern is quantization of particle dynamics in $\Milne_C$.
It means that we do not intend to present the quantization scheme
which is, e.g. boost-invariant.  Quantum theories of a particle in
$\Milne_C$ and $\Milne$ are different because the phase spaces of
both systems have different topologies\footnote{We put the emphasis
on the topology in a quantization scheme because it has basic
importance.}.

Before we begin quantization, it is advantageous  to redefine the
algebra (\ref{com6}). It is known (see
\cite{Piechocki:2003hh,LP,LP2,Brzezinski:1992gu,Kowalski:1998hx,Gonzalez:1998kj}
and references therein) that in case canonical variables
$(\pi,\beta)$ have the topology $\dR^1 \times \dS^1$, it is necessary
to replace $\beta$ by $U:=\exp(i\beta)$, and replace the Poisson
bracket
\begin{equation}\label{q1} \{\cdot,\cdot\} =
\frac{\partial\cdot}{\partial \pi}\frac{\partial\cdot} {\partial
\beta} - \frac{\partial\cdot}{\partial
\beta}\frac{\partial\cdot}{\partial \pi}
\end{equation} by the  bracket
\begin{equation}\label{q2} <\cdot,\cdot> :=
\Big(\frac{\partial\cdot}{\partial
\pi}\frac{\partial\cdot}{\partial U}
    - \frac{\partial\cdot}{\partial U}\frac{\partial\cdot}{\partial
    \pi}\Big)U = \{\cdot,\cdot\}U .
\end{equation}
So, in particular one gets $\:<\pi,U>=U~$, instead of
$~\{\pi,\beta\}=1$.

\subsection{Quantization corresponding to the continuous symmetry case}

Applying the  redefinition (\ref{q2}) to the algebra (\ref{com6})
leads to
\begin{equation}\label{q3} \langle c_1,U_2\rangle
=U_2,~~~~~ \langle c_1,U_3\rangle = U_3,~~~~~ \langle
U_2,U_3\rangle = 0 ,
\end{equation}
where $U_2:=\exp(ic_2)$ and
$U_3:=\exp(ic_3)$, and where the algebra multiplication reads
\begin{equation}\label{q4} \langle\cdot,\cdot\rangle :=
\Big(\frac{\partial\cdot}{\partial
c_1}\frac{\partial\cdot}{\partial U_2} - \frac{\partial\cdot}
{\partial U_2}\frac{\partial\cdot}{\partial c_1}\Big)U_2 +
\Big(\frac{\partial\cdot}{\partial
c_1}\frac{\partial\cdot}{\partial U_3} - \frac{\partial\cdot}
{\partial U_3}\frac{\partial\cdot}{\partial c_1}\Big) U_3 .
\end{equation}
One may verify that (\ref{q4}) defines the Lie
multiplication.

Now, let us quantize the algebra (\ref{q3}). To begin with, we define
the mappings
\begin{equation}\label{q5}
c_1 \rightarrow \hat{c_1}\psi(\beta)\varphi(\alpha):=
-i\frac{d}{d\beta}\psi(\beta)\varphi(\alpha),
\end{equation}
\begin{equation}\label{qq5}
U_2\rightarrow\hat{U_2}\psi(\beta)\varphi(\alpha)
:=e^{i\beta}\psi(\beta)\varphi(\alpha),~~~~~~  U_3\rightarrow
\hat{U_3}\psi(\beta)\varphi(\alpha):=e^{i\beta}\psi(\beta)
e^{i\alpha}\varphi(\alpha),
\end{equation}
where $0\leq\beta,\alpha <2\pi$. The operators $\hat{c_1}, \hat{U_2}$
and $\hat{U_3}$ act on the space $\Omega_\lambda\otimes\Ola$, where
$\Omega_\lambda,~0\leq\lambda < 2\pi,$ is a dense subspace of $L^2
(\dS^1)$ defined as follows
\begin{equation}\label{q6}
\Omega_\lambda = \{\psi\in L^2(\dS^1)~|~ \psi\in C^{\infty}[0,2\pi],
~ \psi^{(n)}(2\pi)=e^{i\lambda}\psi^{(n)}(0),~~~ n=0,1,2,\dots\}.
\end{equation}
The space $\Ola$ may be chosen to have  more general form than
$\Omega_\lambda$. For simplicity, we assume that it is defined by
(\ref{q6}) as well. However, we do not require that $\check{\lambda}
= \lambda$, which means that the resulting representation may be
labelled by $\check{\lambda}$ and $\lambda$ independently.

The space $\Omega_\lambda\otimes\Ola$ is dense in
$L^2(\dS^1\otimes\dS^1)$, so the unbounded operator $\hat{c_1}$ is
well defined. The operators $\hat{U_2}$ and $\hat{U_3}$ are well
defined on the entire Hilbert space $L^2(\dS^1\otimes\dS^1)$, since
they are unitary, hence bounded. It is clear that
$\Omega_\lambda\otimes\Ola$ is a common invariant domain for all
three operators (\ref{q5}) and their products.

One may easily verify that
\begin{equation}\label{q7}
[\hat{c_1},\hat{U_2}]=
\widehat{<c_1,U_2>},~~~~~[\hat{c_1},\hat{U_3}]=
\widehat{<c_1,U_3>},~~~~~[\hat{U_2},\hat{U_3}]=
\widehat{<U_2,U_3>},
\end{equation}
($[\cdot,\cdot]$ denotes commutator), which shows that the mapping
defined by (\ref{q5}) and (\ref{qq5}) is a homomorphism.

The operator $\hat{c_1}$ is symmetric on $\Omega_\lambda\otimes\Ola$,
due to the boundary properties of $\psi\in\Omega_\lambda$. It is
straightforward to show that $\hat{c_1}$ is self-adjoint  by solving
the deficiency indices equation \cite{RS} for the adjoint
$\hat{c_1}^*$ of $\hat{c_1}$ (for more details see Appendix A of
\cite{WP}).

The space $\Omega_\lambda$ may be spanned by the set of orthonormal
eigenfunctions of the operator $\hat{c_1}$ with reduced domain from
$\Omega_\lambda\otimes\Ola$ to $\Omega_\lambda$, which are easily
found to be
\begin{equation}\label{q8}
f_{m,\lambda}(\beta):= (2\pi)^{-1/2}\exp{i\beta
(m+\lambda/2\pi}),~~~~~~m=0,\pm 1,\pm 2,\ldots
\end{equation}
The space $\Ola$ may be also spanned by the set of functions of the
form (\ref{q8}).

We conclude that the mapping defined by (\ref{q5}) and (\ref{qq5})
leads to the self-adjoint representation of (\ref{q3}).

\subsection{Quantization corresponding to the continuous and discrete symmetries case}

Making use of the method presented in preceding subsection we
redefine the algebra (\ref{com8}) to the form
\begin{equation}\label{q9}
\langle c_1,U_2\rangle =U_2,~~~~~ \langle
c_1,U_3\rangle =\varepsilon U_3, ~~~~~ \langle U_2,U_3\rangle = 0,
\end{equation}
where $\var=\pm 1$. We quantize the algebra (\ref{q9}) by the mapping
\begin{equation}\label{h5}
c_1 \rightarrow \hat{c_1}\psi(\beta)f_\var\varphi(\alpha):=
-i\frac{d}{d\beta}\psi(\beta)f_\var\varphi(\alpha),~~~~~~
U_2\rightarrow\hat{U_2}\psi(\beta)f_\var\varphi(\alpha)
:=e^{i\beta}\psi(\beta)f_\var\varphi(\alpha),
\end{equation}
\begin{equation}\label{hh5}
U_3\rightarrow
\hat{U_3}\psi(\beta)f_\var\varphi(\alpha):=e^{i\beta\hat{\var}}
e^{i\alpha}\psi(\beta)f_\var\varphi(\alpha):=
e^{i\beta\var}\psi(\beta)f_\var e^{i\alpha}\varphi(\alpha),
\end{equation}
where $\hat{\var}$ is the operator acting on the two-dimensional
Hilbert space $E$ spanned by the eigenstates $f_\var$ defined by
\begin{equation}\label{q11}
\hat{\var}f_\var =\var f_\var.
\end{equation}
It is easy to check that
\begin{equation}\label{q20}
 [\hat{c_1},\hat{U_2}] =\hat{U_2},~~~~~
[\hat{c_1},\hat{U_3}] =\hat{\varepsilon} \hat{U_3}, ~~~~~
[\hat{U_2},\hat{U_3}] = 0 .
\end{equation}
The domain space of operators  (\ref{h5}) and (\ref{hh5}) is defined
to be the space $\Omega_\lambda\otimes E\otimes\Ola~$. It is evident
that $\hat{\varepsilon}$ commutes with all operators, so the algebra
(\ref{q20}) is well defined. It is easy to check (applying results of
preceding subsection) that the representation is self-adjoint.

\subsection{Quantization in case the system consists of two almost independent parts}

In the last case, the only connection between dynamics in
$\sing_\down$ and $\sing_\up$ is that a particle assumed to exist
in $\sing_\down$, can propagate through the singularity into
$\sing_\up$. It is clear that now  quantization of the system may
be expressed in terms of quantizations done separately in
$\sing_\down$ and $\sing_\up$. To be specific, we carry out the
reasoning for $\sing_\down$:

\noindent The phase space has topology $\Gamma_\down
=\dR^1\times\dS^1$ and the algebra of observables read
\begin{equation}\label{q12}
\langle c_1,U_2\rangle =U_2.
\end{equation}
Quantization of (\ref{q12}) immediately gives
\begin{equation}\label{q13}
    c_1 \rightarrow \hat{c_1}\psi(\beta):=
  -i\frac{d}{d\beta}\psi(\beta),~~~~
  U_2 \rightarrow\hat{U_2}\psi(\beta):=
  e^{i\beta}\psi(\beta),~~~~~~\psi\in\Omega_\lambda,
\end{equation}
which leads to
\begin{equation}\label{q51}
[\hat{c_1},\hat{U_2}]= \widehat{<c_1,U_2>}= \hat{U_2}.
\end{equation}

It is obvious that the same reasoning applies to a particle in
$\sing_\up$.

At this stage we can present quantization of the entire system
having phase space with topology $\Gamma_C :=
\Gamma_\down\times\Gamma_\up$. The algebra of classical
observables reads
\begin{equation}\label{q14}
\langle c_1,U_2\rangle =U_2,~~~~\langle c_4,U_3\rangle =U_3,
\end{equation}
with all other possible Lie brackets equal to zero.

\noindent Quantization of the algebra (\ref{q14})  is defined by
\begin{equation}
c_1 \rightarrow \hat{c_1}\psi(\beta)\varphi(\alpha):=
  -i\frac{d}{d\beta}\psi(\beta)\varphi(\alpha),
~~~~U_2 \rightarrow\hat{U_2}\psi(\beta)\varphi(\alpha):=
e^{i\beta}\psi(\beta)\varphi(\alpha),
\end{equation}
\begin{equation}
c_4 \rightarrow \hat{c_4}\psi(\beta)\varphi(\alpha):=
\psi(\beta)\big(-i\frac{d}{d\alpha}\varphi(\alpha)\big), ~~~~
U_3\rightarrow\hat{U_3}\psi(\beta)\varphi(\alpha):=
  \psi(\beta)e^{i\alpha}\varphi(\alpha),
\end{equation}
where the domain of the operators $\:\hat{c_1}, \hat{c_4},
\hat{U_2}\:$ and $\:\hat{U_3}\:$ is $~\Omega_\lambda\otimes\Ola$.

\noindent It is evident that presented representation is
self-adjoint.

\subsection{Time-reversal invariance}

\noindent The system of a test particle in the Milne space is a
non-dissipative one. Thus, its theory should be invariant with
respect to time-reversal transformation $T$. The imposition of this
symmetry upon the quantum system, corresponding to the classical one
enjoying such an invariance, may reduce the ambiguity of quantization
procedure commonly associated with any quantization method
\cite{AST}.

In our case the ambiguity is connected with the freedom in the choice
of $\lambda$. Since $0\leq\lambda<2\pi$, there are infinite number of
unitarily non-equivalent representations for the algebras of
observables considered in the preceding subsections.  One may  reduce
this ambiguity  following the method of the imposition of
$T$-invariance used for particle dynamics in de Sitter's space.
However, imposition of the rotational invariance on the space of
trajectories makes the definition of time-reversal invariance
meaningless in cases considered in subsections A, B and C of section
III. The $T$-invariance may be imposed only on the dynamics
considered in the subsection D. The first step of quantization for
this case is specified by Eqs. (\ref{q12}) and (\ref{q13}). The
imposition of the $T$-invariance upon the system may be achieved by
the requirement of the time-reversal invariance of the algebra
(\ref{q51}). Formally, the algebra is $\hat{T}$-invariant since
\begin{equation}\label{t1}
    \hat{T}\hat{c_1}\hat{T}^{-1} = -
    \hat{c_1},~~~~~\hat{T}\hat{U_2}\hat{T}^{-1}=\hat{U_2}^{-1},
\end{equation}
where $\hat{T}$ denotes an anti-unitary operator corresponding to the
transformation $T$. The first equation in (\ref{t1}) results from the
correspondence principle between classical and quantum physics,
because $c_1$ has interpretation of momentum of a particle. The
assumed form of $\hat{U_2}$ and  anti-unitarity of $\hat{T}$ lead to
the second equation in (\ref{t1}). The formal reasoning at the level
of operators should be completed by the corresponding one at the
level of the domain space $\Omega_\lambda$ of the algebra
(\ref{q51}). Following step-by-step the method of the imposition of
the $T$-invariance upon dynamics of a test particle in de Sitter's
space, presented in Sec.(4.3) of \cite{Piechocki:2003hh}, leads to
the result that the range of the parameter $\lambda$ must be
restricted to the two values: $\lambda=0$ and $\lambda=\pi$.

Now, let us take into account that quantum theory is expected to be
more fundamental than its classical counterpart (if the latter
exists). In the context of the time-reversal invariance it means that
$\hat{T}$-invariance may be treated to be more fundamental than
$T$-invariance. Applying this idea to quantum particle in the Milne
space, we may ignore the lack of $T$-invariance of classical dynamics
considered in subsections A, B and C. In these cases we propose to
mean by the time-reversal invariance the $\hat{T}$-invariance only.
It may be realized  by the requirement of $\hat{T}$-invariance of the
corresponding algebras. For instance, the algebra (\ref{q20}) is
formally $\hat{T}$-invariant if the observables transform as follows
\begin{equation}\label{t2}
\hat{T}\hat{c_1}\hat{T}^{-1} =-\hat{c_1},
~~~~~\hat{T}\hat{U_2}\hat{T}^{-1}=\hat{U_2}^{-1},
~~~~~\hat{T}\hat{U_3}\hat{T}^{-1}=\hat{U_3}^{-1},
~~~~\hat{T}\hat{\var}\hat{T}^{-1} =\hat{\var}~.
\end{equation}
We require the first equation of (\ref{t2}) to hold. All other
equations in (\ref{t2}) result from the functional forms of
$\hat{U_2}$, $\hat{U_3}$ and $\hat{\var}$, and the anti-unitarity of
$\hat{T}$. These analysis should be completed by the corresponding
one at the level of the the domain space $\Omega_\lambda\otimes
E\otimes\Ola~$ of the algebra (\ref{q20}), but we do not enter into
such details.

The imposition of $\hat{T}$-invariance not only meets the expectation
that a system with no dissipation of energy should have this
property, but also helps to reduce the quantization ambiguity as it
was demonstrated in the simplest case (It is clear that three other
cases enjoy this reduction too.).

\section{Relation to the FLRW comologies}

 It appears that our model of a particle in compactified 2D Milne
space is only a toy model. In fact it may be linked to some cosmology
models obtained by compactification to hyperbolic scalar target
spaces of some higher-dimensional string/M theories
\cite{Russo:2004am,Bergshoeff:2005cp,Bergshoeff:2005bt}. It is very
interesting that some cosmological solutions of these effective field
theories can be interpreted\footnote{Due to the construction based on
the Maupertuis-Jacobi principle of classical mechanics
\cite{Russo:2004am}.} in terms of time-like straight-line geodesics
in higher-dimensional noncompactified Milne space $\Milne$. It turns
out that points at which these straight lines meet the Milne horizon
may correspond to cosmological singularities of the original
effective field theories.

To be specific, let us take a straight-line time-like geodesic in the
3D moduli space of flat FLRW universe, $\Milne$, of the above field
theory presented in Fig. 1 of \cite{Russo:2004am}, and  let us make
the mapping of this  straight-line  into our 2D $\Milne_C$ space. It
is not difficult to see that the image of its part which belongs to
the past-Milne, is a spiral geodesic in the lover cone of $\Milne_C$.
The part which lies in the future-Milne, maps onto a spiral geodesic
of the upper cone of $\Milne_C$. The resulting spiral geodesic in
$\Milne_C$ belong to the type of geodesics  considered in subsection
D. The big-crunch and big-bang represented in $\Milne$ by two
intersection points of the straight-line with the Milne horizon are
mapped into the vertex of $\Milne_C$, i.e. into the
big-crunch/big-bang singularity of our model spacetime.   Therefore,
at the classical level the big-crunch/big-bang singularity of
$\Milne_C$ space may correspond to big-crunch and big-bang
singularities  of some effective higher-dimensional field theories.

Next  step is investigation of this analogy  at the quantum level. In
paper \cite{Russo:2004am} the part of time-like straight-line
trajectory which is in the Rindler space, i.e. `non-Milne' region of
space,  is interpreted as forbidden, because the scale factor,
$\eta$, of the corresponding FLRW becomes complex there. The
interpretation  is that in that `hidden' region quantum effects
should be taken into account. The quantization is carried out by
imposition of some suitable phase space constraint\footnote{The
constraint is connected with the Friedmann constraint of  FLRW
dynamics which translates into the mass-shell constraint of particle
dynamics in the moduli space of FLRW universe.} as an operator
constraint on a Hilbert space. Solution to this equation defines the
wave function of the universe which is analytic in $\eta$. The final
conclusion is  that a collapsing universe can pass the  classically
forbidden region, owing to the quantum mechanical tunnelling, into a
region where it becomes an expanding universe \cite{Russo:2004am}. In
our case quantum description of a particle in $\Milne_C$,
corresponding to the dynamics presented in the subsection D, is
mathematically well defined. It is clear that there must exist some
relation between both results. The problem is that our quantization
method is quite different from the method applied in
\cite{Russo:2004am}. Finding explicit relation between both quantum
models needs further analysis \cite{PWN}.

Comparison of our results with the results of
\cite{Bergshoeff:2005cp,Bergshoeff:2005bt} will become possible after
we generalize the analysis of particle dynamics from compactified
Milne  to compactified Misner space. The latter includes the
compactified Rindler space. By extending  reasoning of the previous
paragraph, we  expect that particle dynamics in the compactified
Rindler space may be related with the instanton phase of particle
propagation in the moduli space of FLRW universes . As the result,
the full cosmology/instanton solutions presented in
\cite{Bergshoeff:2005cp,Bergshoeff:2005bt} may be linked to the
dynamics of a particle in the compactified 2D Misner space. We
postpone investigation of this possibility to our next papers.

\section{Conclusions}

Finding specific model(s) of a quantum particle in the compactified
Milne space is our main result.  We have analysed four ways of
particle's transition across the singularity, but more cases are
possible.  It is so  because at the singularity the equations
defining classical dynamics are not well defined.  It is the direct
consequence of the fact that at the singularity the space dimension
disappears, which causes that the Cauchy problem for time-like
geodesics is not well defined. In case there is no clear reason to
choose specific transition across the singularity, it  acts as
`generator' of uncertainty in the propagation of a particle from the
pre- to post-singularity era.

Extension of our model to higher dimensional compactified Milne space
may be carried out (as it was done in case of de Sitter space
\cite{Jorjadze:1999wb}) to make it more realistic, but we do not
expect that the main conclusion would be changed owing to the Cauchy
problem at the singularity which could not be avoided. However, such
generalization should be done due to the connection of our results to
cosmology models of higher dimensional effective field
theories\footnote{In general, moduli spaces of these theories are
higher dimensional.} considered in
\cite{Russo:2004am,Bergshoeff:2005cp,Bergshoeff:2005bt}.

The  quantum theory depends on the assumptions one makes for the
passage of a particle through  the singularity. This way, however,
one may put forward some hypothesis concerning its nature.  Such
flexibility does not occur in case of de Sitter space, owing to the
uniqueness of particle dynamics \cite{WP,Piechocki:2003hh}.

 It is amazing that time-like geodesics in $\Milne_C$ may have
interpretation in terms of cosmological solutions of some
sophisticated higher dimensional field theories. This connection
deserves further investigation especially at the quantum level to
reveal the nature of the cosmic singularity \cite{PWN}.

Our analysis are based on the assumption that a classical particle is
able to pass the singularity. Justification for such assumption is
that we consider a test particle. Physical particle might collapse
into a black hole at the singularity, modify the spacetime there, or
both. We have also ignored the effect of particle's own gravitational
field on its motion \cite{Poisson:2004gg}.  Some modelling   of these
effects \cite{PWN} may be carried out by considering particle
dynamics in a spacetime which regularizes  the space $\Milne_C$.

Our model concerns point-like objects. Next natural step would be
examination of dynamics of extended objects like strings or
membranes. According to string/M theory (see, e.g. \cite{JP}), they
are more elementary than  point particles. It was recently shown (see
\cite{Durin:2005ix,Pioline:2003bs,Turok:2004gb,Nekrasov:2002kf} and
references therein) that a  test string in the zero-mode state
twisted around the shrinking dimension propagates smoothly and
uniquely across the Milne space singularity. It is interesting that
strings in such states do not suffer from the blue-shift effect
specific for a point particle. However, as it was pointed out in
\cite{Pawlowski:2005bs}, understanding of propagation of a string in
the zero-mode state is not the end of the story. For drawing firm
conclusions  about the physics of the problem one should also examine
the non-zero string modes and go beyond the semi-classical
approximation.

\appendix

\section{}

In this section we present solutions to the equations (\ref{var}).
For the Lagrangian (\ref{lag}) the equations read
\begin{equation}\label{ruch} \frac{d}{d\tau}\bigg(\frac{m
t^2\dot{\theta}}{e}\bigg)=0,~~~~~~
\ddot{t}-\bigg(\frac{\dot{e}}{e}\bigg)\dot{t}+\dot{\theta}^2t=0,~~~~~~
e^2=\dot{t}^2-t^2\dot{\theta}^2 .
\end{equation}
From the first and third equations of  (\ref{ruch}) it is easily seen
that
\begin{equation}\label{r1}
m t^2 \dot{\theta} =e\: c_1~~~~~~\textrm{and}~~~~~~~ t^2
\dot{\theta}^2 = \dot{t}^2 - e^2, \end{equation} respectively,
where $c_1 \in\dR$. Combining equations of (\ref{r1}) we see at
once that \begin{equation}\label{r2} e^2 = t^2\dot{t}^2/((c_1/m)^2
+t^2). \end{equation} Making use of (\ref{r2}) to rewrite the
identity  $\dot{e}/e = \frac{d}{d\tau}e^2/2e^2\:$ ($e\neq 0$ for
time-like geodesics) and substituting the resulting expression
into (\ref{ruch}) yields \begin{equation}\label{rowruch}
\textrm{sgn}(\dot{\theta})=\textrm{sgn}(e)\:\textrm{sgn}(c_1),~~~~~~
\dot{\theta}^2=\frac{c_{1}^2\dot{t}^2}{m^2 t^4+c_{1}^2t^2},~~~~~~
e^2=\dot{t}^2-t^2\dot{\theta}^2. \end{equation} (In what follows
we choose $\textrm{sgn}(e)=1$ to be specific.)

\noindent Now, we will find  solution to (\ref{rowruch}). Since
the action (\ref{action}) is reparametrization invariant, the
equations (\ref{rowruch}) have  this property too.  The mapping
$\tau\mapsto\overline{\tau}$ leads to \begin{equation}
e^2\mapsto\bigg(\frac{d\tau}{d\overline{\tau}}\bigg)e^2\bigg(\frac
{d\tau}{d\overline{\tau}}\bigg),
~~~~~~\dot{\theta}\mapsto\bigg(\frac{d\tau}{d\overline{\tau}}\bigg)\dot{\theta},
~~~~~~\dot{t}\mapsto\bigg(\frac{d\tau}{d\overline{\tau}}\bigg)\dot{t}
\end{equation}
It means that we can arbitrarily choose either $e$ or $\dot{t}$.
Since we consider time-like geodesics,  we cannot choose an
arbitrary $\dot{\theta}$ (there exist solutions to (\ref{rowruch})
with $\dot{\theta}=0$ and we are unable to assign to them other
values). Let us choose \begin{equation}\label{repar}
\dot{t}=1,~~~~~\textrm{or equivalently}~~~~~\tau:=t+C.
\end{equation} (For simplicity we put $C=0$.) The gauge
(\ref{repar}) leads to \begin{equation}\label{sol}
\bigg(\frac{d\theta}{dt}\bigg)^2=\frac{c_{1}^2}{m^2
t^4+c_{1}^2t^2},~~~~~~ e^2=\frac{t^2}{t^2+(c_1/m)^2}.
\end{equation} The solution of (\ref{sol}) reads
\begin{equation}\label{soll} \theta(t)=
-\int\frac{d(\frac{c_{1}}{mt})} {\sqrt{1+(\frac{c_{1}}{mt})^2}}=
-\textrm{arsinh} \bigg(\frac{c_{1}}{mt}\bigg)+c_2,~~~~~~0 \leq c_2
<2\pi. \end{equation} (Note that $\textrm{sgn}(c_1)=1$, due to the
choice $\textrm{sgn}(e)=1$ done after (\ref{rowruch})).

Now we determine $p_t$ and $p_\theta\:$. Applying (\ref{r1}) and
(\ref{r2}) we get
\begin{equation}\label{r3}
p_t =\partial L/\partial\dot{t}= -m\frac{\dot{t}}{e}=m\:
\sqrt{1+\bigg(\frac{c_1}{mt}\bigg)^2}, ~~~~~~~~p_\theta= \partial
L/\partial\dot{\theta}=m \frac{\dot{\theta}t^2}{e}= c_1 .
\end{equation} Finally, we rewrite  the solution (\ref{soll}) in
terms of the dynamical integrals (\ref{in1})-(\ref{in3}). Let us
notice that for $c_1 =0\:$ we have \begin{equation}\label{pom}
\theta =c_2~~~~~~\textrm{and}~~~~~~I_2/I_1 =\tanh\theta.
\end{equation} Hence for $\:c_1 =0\:$ we get $\:\theta =
\tanh^{-1}(I_2/I_1)$. Thus, the solution reads
\begin{equation}\label{r4} \theta(t)= -\sinh^{-1}
\bigg(\frac{I_3}{mt}\bigg)
+\tanh^{-1}\bigg(\frac{I_{2}}{I_{1}}\bigg).
\end{equation}
We conclude that $\:c_1\:$ and $\:c_2\:$ parametrizing geodesics
have the interpretation of particle observables.

Using the solution (\ref{soll}), we can rewrite the dynamical
integrals in terms of $c_1$ and $c_2$ as follows
\begin{equation}
I_1=m\cosh(c_2),~~~~~~I_2=m\sinh(c_2),~~~~~~I_3=c_1.
\end{equation}
Introducing the Poisson bracket by
\begin{equation}\label{ccom}
\{\cdot,\cdot\} := \frac{\partial\cdot}{\partial
c_1}\frac{\partial\cdot} {\partial c_2} -
\frac{\partial\cdot}{\partial c_2}\frac{\partial\cdot} {\partial
c_1},
\end{equation} one may easily verify that
\begin{equation}\label{rel1} \{I_1,I_2\}=0,~~~~~\{I_3,I_2\}=I_1,
~~~~~\{I_3,I_1\}=I_2.
\end{equation}
Thus, the algebras (\ref{rel1}), (\ref{coom}) and (\ref{com1}) are
isomorphic.  Eqs. (\ref{rel1}) characterize the local symmetry of
the system for either $t<0$ or $t>0$. Owing to the obvious
relation $\{c_1,c_2\}=1$, the variables $c_1$ and $c_2$ are
canonical.

\begin{acknowledgments}
We would like to thank Jean-Pierre Gazeau, Marek Paw\l owski, Boris
Pioline, Micha\l\ Spali\'nski and Neil Turok for helpful discussions,
and to the anonymous referees for constructive criticisms. One of us
(WP) would like to thank the European Network of Theoretical
Astroparticle Physics ILIAS/N6 under contract number
RII3-CT-2004-506222 for partial financial support.
\end{acknowledgments}

\end{document}